\documentclass[final]{aipproc}
\layoutstyle{6x9}
\begin{document}

\title{Collective Atomic Recoil Lasing and Superradiant Rayleigh Scattering in a high-Q ring
cavity}


\author{Sebastian Slama}{}

\author{Gordon Krenz}{}

\author{Simone Bux}{}

\author{Claus Zimmermann}{}

\author{Philippe W. Courteille}{
 address={Physikalisches Institut, Eberhard-Karls-Universität
Tübingen, Auf der Morgenstelle 14, D-72076 Tübingen, Germany}}

\begin{abstract}
Cold atoms in optical high-Q cavities are an ideal model system
for long-range interacting particles. The position of two
arbitrary atoms is, independent on their distance, coupled by the
back-scattering of photons within the cavity. This mutual coupling
can lead to collective instability and self-organization of a
cloud of cold atoms interacting with the cavity fields. This
phenomenon (CARL, i.e. Collective Atomic Recoil Lasing) has been
discussed theoretically for years, but was observed only recently
in our lab. The CARL-effect is closely linked to superradiant
Rayleigh scattering, which has been intensely studied with
Bose-Einstein condensates in free space. By adding a resonator the
coherence time of the system, in which the instability occurs, can
be strongly enhanced. This enables us to observe cavity-enhanced
superradiance with both Bose-Einstein condensates and thermal
clouds and allows us to close the discussion about the role of
quantum statistics in superradiant scattering.
\end{abstract}

\maketitle

\section{Introduction}
An astonishing phenomenon is reported from populations of
fireflies \cite{Buck66}. Individual fireflies are emitting light
in a randomly flashing way. However, if the population exceeds a
certain number of individuals, the uncorrelated flashing of the
single flies synchronizes such that a large part of the population
flashes in phase with each other and 'seems to agree' on a common
flashing frequency. This principle of self-synchronization can be
found in many other biological, social, chemical and physical
systems, as for example in the frequency-locking of laser arrays
or in the walking behavior of people on bridges \cite{Strogatz01,
Strogatz05}. All of these systems can be described in a very
general model, the Kuramoto model, which is based on a collection
of weakly coupled oscillators \cite{Kuramoto84}.\\

The same model applies to clouds of cold atoms, if the individual
atoms experience an interaction. Such an interaction can for
example be generated by the scattering of photons. If the
interaction is weak, atoms scatters photons individually and are
not influenced by the scattering of other atoms. This is even true
for Bose-Einstein condensates (BEC), where all atoms are
delocalized over the same region of space. If the atoms do not
show long-range order, such individual scattering manifests in a
scattered light intensity which is proportional to the number of
scatterers. In contrast, if the interaction is strong enough, the
individual scattering events can be synchronized such that all
atoms contribute in a cooperative way. Scattering then evolves as
a global dynamics of the whole cloud, with the scattered intensity
being enhanced by the phase coherent emission of light fields and
depending on the number of scatterers with a power law larger than
one. A prominent example for this behavior is Dicke superradiance
\cite{Dicke54}: a dense cloud of electronically excited atoms
synchronizes its spontaneous emission and emits a light pulse with
an intensity which depends quadratically on the number of atoms.
However, we must bear in mind that similar dependencies can also
arise due to an externally imposed long-range order of the cloud
rather than to self-synchronization. This is for example true for
Bragg-scattering from atomic clouds which are periodically
arranged by means of an optical lattice \cite{Birkl95,
Weidemüller95, Slama05, Slama06}. Interestingly, long-range order
can arise due to the self-synchronization process itself. In this
case we talk about self-organization of structures. Again there is
a tremendous amount of examples of self-organizing real systems of
which just one is the formation of natural landform
patterns such as meandering rivers or sand dunes \cite{Werner99}.\\

In this paper we are dealing with two types of experiments with
ultracold atoms, which are coupled by the scattering of photons,
namely superradiant Rayleigh Scattering (SRyS) on one hand
\cite{Inouye99, Kozuma99} and Collective Atomic Recoil Lasing
(CARL) on the other \cite{Slama07a, Slama07b}. Both experiments
show self-organization leading to the spontaneous formation of a
density grating of atoms.  We will present our measurements of
CARL in different regimes by which we will show that the two
experiments are instrinsically connected with each other.\\

\section{CARL and SRyS}
CARL has been proposed for atomic gases which are illuminated by a
strong optical pump field \cite{Bonifacio94, Bonifacio95}. The
atoms scatter photons from the pump field into an unpumped probe
field. The acceleration due to photonic recoil, experienced by the
atoms, depends on the atomic position in such a way that the atoms
are arranged in a density grating, which fulfills the Bragg
condition and in turn enhances the scattering. This effect is
particularly strong, if pump- and probe light fields are modes of
an optical high-Q ring cavity and, up to now,
has only been observable in cavities.\\

SRyS on the other hand has been observed in free space. The atoms
scatter light from a short pump pulse into a probe light field and
are accelerated in units of the recoil momentum $p_r=2\hbar k$,
with $k$ the wavevector. This leads to the occupation of different
momentum states, which by matter-wave interference build up a
density grating and in turn enhance the scattering. Due to the
fact that, at the beginning, SRyS could only be observed with
BECs, and although in 2005 superradiant Raman scattering was
observed with atoms slightly hotter than the critical temperature
for Bose-Einstein condensation \cite{Yoshikawa05}, some
discussions about the role of quantum statistics and the
importance of Bose-enhancement for SRyS arose\cite{Ketterle01,
Moore01}. From our present point of view, the prediction that the
relevant phenomenon for SRyS is not the quantum state of the atoms
but their cooperative behavior \cite{Inouye99}, was already
confirmed by the first observation of CARL \cite{Kruse03} with
$100~\mu\textrm{K}$ cold atoms, which is two orders of magnitude
hotter than the condensation temperature. But it took us until now
when we observed CARL in a regime with superradiance-like dynamics
and directly measured the dependence on the temperature, that we
fully understood the connection between CARL and SRyS and could
generalize our former results
to superradiance.\\

What CARL and SRyS have in common is the gain mechanism
\cite{Piovella97}, which in both cases is based on a positive
feedback between a probe light field and an atomic density
grating. The gain, which is given by the growth rate of the probe
light field, depends on the frequency of the light field
$G(\omega)$ and has a certain gain bandwidth
$\delta\omega_{\textrm{G}}$. The analogy shows even up as formal
identity of the gain in both cases \cite{Piovella01, Robb05,
Slama07c}. Due to this formal concordance we might wonder, why
CARL works with rather hot thermal atoms, while SRyS is restricted
to the ultracold regime. The answer lies in the fact that the
self-amplification works only within the coherence time of the
system. The difference between the effects is, how the coherence
of the system is preserved. In the case of SRyS the coherence time
is given by the stability of the relative phase between different
atomic matter waves. If the relative phase is smeared out, the
density grating vanishes and the superradiant emission of light is
stopped. For this reason SRyS is very sensitive to temperature
effects, because a defined phase can only be attributed to the
matter waves if the atomic Doppler broadening is much smaller than
the momentum imparted by a single photonic recoil. In the case of
CARL the situation is changed by the presence of the cavity, which
leads to a very long life-time of both the pump- and the probe
light field on the order of several $\mu\textrm{s}$, which is
given by the cavity linewidth $\kappa_{\textrm{c}}$. The coherence
can thus be preserved as the relative phase of the cavity
light-fields, which is independent of the atomic motion. The atoms
are then forced to maintain in the density grating by the optical
lattice in the cavity, even if they are rather hot. Temperature
effects on CARL do exist and we examine those within this paper,
but these are much less dramatic than in the experiments dealing
with SRyS.\\

\section{CARL in various regimes}
            \begin{figure}\label{fig:regime}
            \includegraphics[height=.3\textheight]{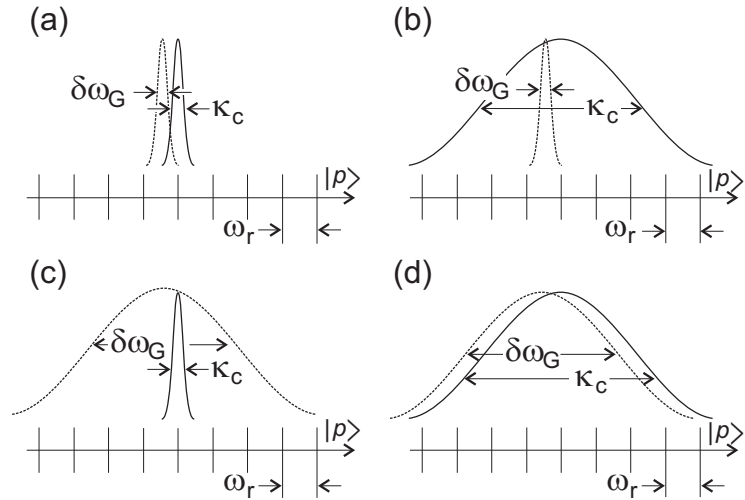}
            \caption{Schematic illustration of the four CARL regimes. (a) Quantum good-cavity regime,
            (b) quantum bad-cavity regime, (c) semiclassical good-cavity regime and (d) semiclassical bad-cavity regime.}
            \end{figure}
CARL may occur in four different regimes with each of them showing
a characteristic dynamics of the atoms (Fig.\ref{fig:regime}).
Until now, all CARL-experiments \cite{Kruse03, Cube04} had been
performed in the good cavity regimes, whereas all experiments
dealing with SRyS \cite{Inouye99, Kozuma99, Yoshikawa05} had taken
place very far in the bad-cavity regimes. In our latest
experiments \cite{Slama07a, Slama07b} we have been able to examine
the transition between these regimes by changing the finesse of
the cavity. What discerns the good cavity from the bad cavity
regime is the size of the cavity linewidth compared with the
recoil frequency $\omega_{\textrm{r}}=p_r^2/2m\hbar$. The
linewidth determines the density of states inside the cavity and
limits the range of frequencies accessible for the probe light
field. The frequency of the scattered photons is on the other hand
Doppler shifted with respect to the pump light frequency, which we
lock to a cavity resonance. The size of the shift is given by the
momentum of the scattering atoms. For that reason a cavity
linewidth which is smaller than the recoil frequency (which is the
distance between neighboring momentum states in frequency space),
like shown in Fig. \ref{fig:regime} (a) and (c), limits the atomic
dynamics to the momentum state $|p\rangle=|0\rangle$ and its
closest neighbors. This case is called the good cavity regime. If
on the other hand, like in Fig. \ref{fig:regime} (b) and (d), the
cavity linewidth is larger than the recoil frequency, all momentum
states which are lying within the linewidth may participate in the
CARL dynamics. In that sense the difference between good and bad
cavity is the principle frame, in which the dynamics may take
place. But how the dynamics takes place within this frame is
determined by yet another variable, namely the gain bandwidth
$\delta\omega_{\textrm{G}}$. The ratio between the gain bandwidth
and the recoil frequency determines, how many momentum states are
amplified at the same time. If for example the ratio is smaller
than one, Fig. \ref{fig:regime} (a) and (b), which is called the
quantum regime, only two neighboring momentum states are coupled
with each other. This leads to a coherent behavior like in a two
level system, such that at any time the momentum population is
distributed to a maximum value of just two momentum states. If one
the other hand the ratio between the gain bandwidth and the recoil
frequency is larger than one, Fig. \ref{fig:regime} (c) and (d),
which is called the semiclassical regime, several momentum states
are coupled at a time. In this case the initial momentum
distribution, even if only one momentum state was occupied like in
a BEC, is spread over several momentum states by the dynamics. The
occupation of more and more momentum states then leads to a
decreasing bunching of the atoms and consequently to decoherence
of the system. \footnote{While discerning between good cavity, bad
cavity, quantum and semiclassical regime, we have to bear in mind,
that the explained distinction by cavity linewidth and gain
bandwidth is useful for understanding the underlying physics, but
the quantities delimiting the regimes have a complicated
interdependence. It is for example possible, that the dynamics
would be, according to the cavity linewidth, in the good cavity
regime. Nevertheless a very large gain bandwidth may broaden the
spectrum of excited momentum states as in the bad cavity regime. A
thorough description resorts to dimensionless, independent
parameters, which are called CARL parameter $\rho$ and scaled
linewidth $\kappa$ \cite{Piovella01}.}

\section{Equations of motion}
Our system consists of ultracold atoms and Bose-Einstein
condensates interacting with the counterpropagating modes of a
high-Q ring cavity. For a general approach which describes mean
field interactions or quantum statistical effects such as nonlocal
particle correlations, particle fluctuations or entanglement, both
the light and the matter wave modes would have to be treated as
quantized \cite{Moore99, Piovella03}. However, in the case of our
experiments, several simplifications can be made:\\

(1) The detuning of the pump laser frequency to the closest atomic
resonance is on the order of several THz, such that the population
of all electronically excited states is negligible. These states
can then by adiabatically eliminated \cite{Bonifacio95}. (2) For
the same reason the optical density of the atomic cloud is so low,
that propagation effects of light inside the cloud need not to be
considered \cite{Bonifacio97}. (3) In the semiclassical regime,
where our experiments take place, quantum effects, such as
entanglement which may occur in CARL \cite{Piovella03}, are
negligibly small. (4) We are working with high laser powers, such
that the light fields can be treated classically. (5) We are
describing the dynamics of our system only along the cavity axis,
i.e. in one dimension. Transversal effects in CARL exist
\cite{Piovella06}, but are not considered, because they are much
weaker than the axial dynamics. (6) The backaction of the atoms on
the pump light field is neglected, because typical probe light
powers are three orders of magnitude weaker than the pump light
power, which allows us to make a nondepleted pump approximation.
Experimentally the pump laser frequency is phase-locked to a
resonance of the cavity, such that there is a fixed phase-relation
between the pump laser field and the intracavity pump mode field
amplitude $\alpha_+$. The lock compensates also for the shift of
the cavity resonance due to the change of the refractive index
when atoms are loaded into the cavity. We keep our system in the
regime of weak coupling between the cavity modes and the atoms,
where splitting of the cavity resonance into two normal modes,
which was observed in \cite{Klinner06}, is negligible. For these
reasons the phase of the pump mode field can be chosen
arbitrarily, which we do by setting $\alpha_+$ real. Within these
approximations our system is well described by the equations

\begin{eqnarray}\label{Eq1}
    \frac{dp_j}{dt}  &=-2i\hbar kU_0\alpha_+\left(\alpha_-^*e^{2ikz_j}-\alpha_-e^{-2ikz_j}\right)~,\nonumber\\
    \frac{d\alpha_-}{dt}
    &=-(\kappa_c+i\Delta_c)\alpha_--iU_0\alpha_+\sum_{j=1}^Ne^{-2ikz_j}~.
\end{eqnarray}

These equations describe the dynamics of $N$ atoms with momenta
$p_j$ and positions $z_j$ and of the probe light field $\alpha_-$.
The force acting on the atoms is the dipole force, which is
proportional to the single photon light shift $U_0$. The left term
in the lower equation for the probe light field contains losses
with the cavity linewidth $\kappa_c$ and phase shifts which are
proportional to the detuning from the cavity $\Delta_c$ (in our
case is $\Delta_c=0$). The right term represents the scattering of
photons from the pump mode into the probe mode and is proportional
to the bunching of the atoms, which is given by
$b=N^{-1}\sum_{j}e^{-2ikz_j}$~. Radiation pressure and light
scattering from the mirror surfaces lead to observable
perturbations in our experiments and can also be included into the
equations, as shown in \cite{Slama07b}.

\section{Experimental procedure and Measurements}
            \begin{figure}\label{fig:setup}
            \includegraphics[height=.4\textheight]{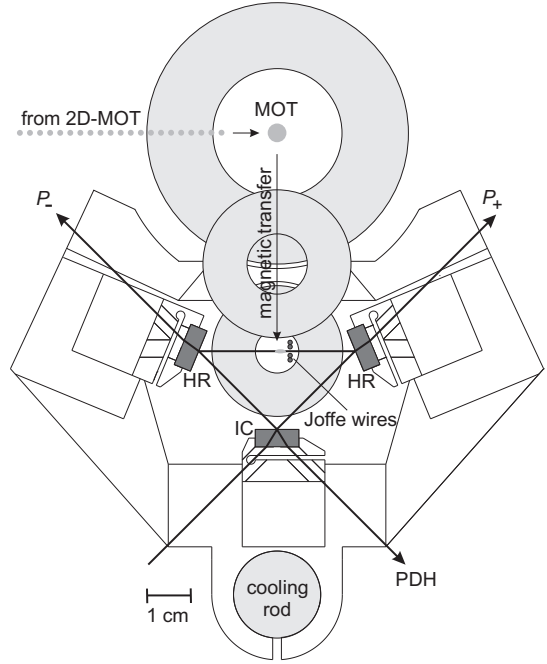}
            \caption{Technical drawing of our setup.}
            \end{figure}
Our setup shown in Fig. \ref{fig:setup} combines the techniques to
produce Bose-Einstein condensates with a high-Q high ring cavity.
All parts including magnetic coils, wires and the cavity are
located inside a vacuum chamber. A cold atomic beam of
$^{87}\textrm{Rb}$ is produced by a two-dimensional
magneto-optical trap (2D-MOT) within a second vacuum chamber and
is directed into the main chamber, where it is recaptured in a
standard MOT. From there the atoms are loaded into a magnetic
trap, which is formed by the same quadrupole coils as the MOT.
Then, the atoms are transferred magnetically via a second into a
5~cm distant third pair of coils and are thereby adiabatically
compressed. The compressed atoms are loaded into a Joffe-Pritchard
type of wire trap produced by the coils and four vertically
directed wires, where they are cooled by forced evaporation with a
microwave frequency. Typical oscillation frequencies in this
harmonic trap are $\omega_r=200~\textrm{Hz}$ and
$\omega_z=50~\textrm{Hz}$ at a magnetic offset field of
$B_0=2~\textrm{G}$ with the longitudinal z-direction pointing
along the cavity axis. We can produce Bose-Einstein condensates
with typically several $10^5$ atoms. For most of the experiments
reported in this paper however, we have been using ultracold
thermal clouds, because quantum degeneracy is unnecessary for
CARL. After the evaporation the atoms are transferred vertically
about 1~mm into the mode volume of the cavity. The cavity consists
of one plane (IC) and two curved (HR) mirrors with a curvature of
$R_c=10~\textrm{cm}$. The round-trip length of the cavity is
8.5~cm, corresponding to a free spectral range of
$\delta_{\textrm{fsr}}=3.5~\textrm{GHz}$. The finesse of the
cavity depends on the polarization of light and adopts values of
$F=87000$ for p-polarization and $F=6500$ for s-polarization.\\
            \begin{figure}\label{fig:timesignals}
            \includegraphics[height=5cm]{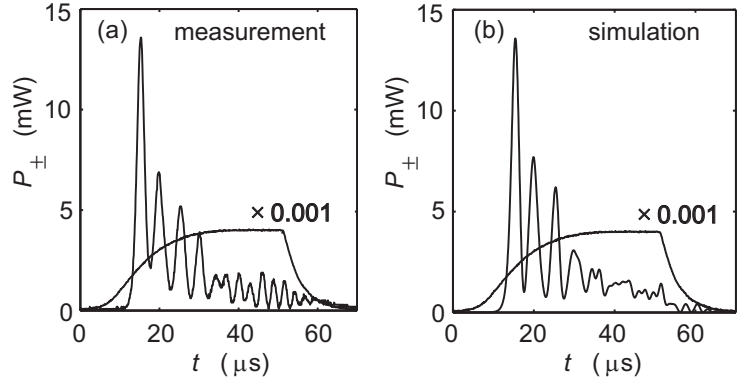}
            \caption{(a) Typical measured time curves of pump and probe light power. In
            this measurement the experimental parameters are $N=1.5\times 10^6$, $P_+=4~\textrm{W}$,
            $\lambda=797.3~\textrm{nm}$, and $F=87000$. For visibility the pump-light power is scaled
            down by a factor of 0.001. (b) Simulation of the CARL-equations. We used the measured rise of the
            pump-light power within the simulation and fitted the experimental parameters in order
            to agree with the measurement. The fitted parameters are in reasonable agreement with the measured ones.}
            \end{figure}
As soon as the atoms are positioned inside the cavity, we switch
on the pump light, which is fed into the cavity via the IC mirror.
The light reflected from the mirror is used to Pound-Drever-Hall
(PDH) stabilize the light frequency to the cavity resonance. Due
to the finite bandwidth of the locking servo is takes
approximately $20~\mu\textrm{s}$ until the pump light field in the
cavity has built. After a time of $t=50~\mu\textrm{s}$ we switch
off the pump light field. In the mean time the CARL dynamics is
starting and a train of probe light pulses like shown in Fig.
\ref{fig:timesignals} is emitted. This characteristic behavior can
be most easily explained within the quantum regime, where the
atomic momenta are distributed only between neighboring states. If
all atoms are in the same momentum state, like in a BEC, light
scattering is suppressed, because the position distribution of a
single momentum state is homogeneous. When however, due to CARL,
more and more atoms are brought into the neighboring momentum
state, matter-wave interference leads to an increasingly developed
density grating of atoms, which enhances the scattering. The
maximum contrast of the grating is reached, when the two
neighboring momentum states are equally populated. This is the
point, when also the probe light power reaches its maximum. The
further dynamics again concentrates the population in one of the
two momentum states, and the density grating and hence the probe
light are vanishing. This dynamics repeats in a self-similar way,
until decoherence smears out the dynamics. This can be seen in
Fig. \ref{fig:timesignals} as decrease of the envelope of the
train of pulses and is due to the increasing spread of the
occupied momentum states in the semiclassical regime. In the
following chapters we will examine the dependence of the probe
light power on certain experimental parameters. In particular we
analyze the power at the first maximum $P_{-,1}$ and the time
interval between the first and the second maximum $\Delta
t_{1,2}$, which is the typical timescale, in which the momentum
changes by one recoil $p_\textrm{r}=2\hbar k$. The parameters we
are changing are the atom number $N$, the intracavity pump light
power $P_+$, the finesse of the cavity $F$ and the atomic
temperature $T$. Dependences on perturbations such as mirror
backscattering are reported in \cite{Slama07b}. Simulation results
which we compare with our measurements are obtained by numerical
integration of Eq. (\ref{Eq1}) with the explicit Euler method,
calculating the trajectories of $N_\textrm{S}=100$ atoms, each
representing $N/N_\textrm{S}$ real atoms.

\subsection{Pump power}
            \begin{figure}\label{fig:pumppower}
            \includegraphics[height=5cm]{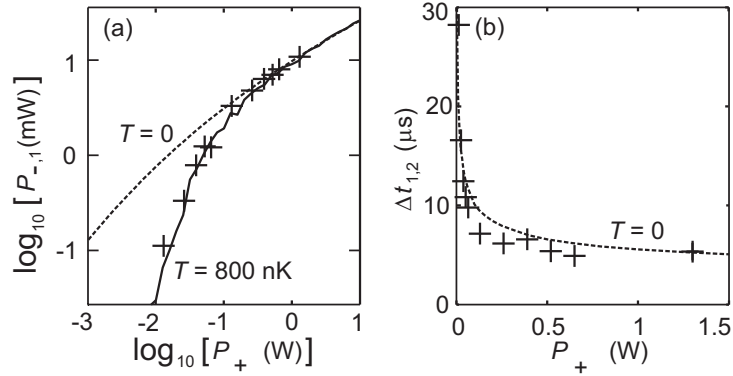}
            \caption{Dependence of (a) the peak probe power $P_{-,1}$ and (b) the time delay between
            the first two peaks $\Delta t_{1,2}$ on the pump power $P_+$. Simulations are done without free
            parameters. Residual fluctuations in the simulation for $T=800~\textrm{nK}$ are due to the statistics
            occuring for a limited simulated atom number of $N_S=100$. The experimental parameters are
            $N=2.4\times 10^6$, $\lambda=796.1~\textrm{nm}$, and
            $F=87000$. The stochastic error of the measurements
            is smaller than the marker size.}
            \end{figure}
The pump light power influences the dynamics of the collective
instability. If the pump power is reduced, also the contrast of
the optical standing wave decreases, which is formed by
interference between pump and probe light field. This weakens the
collective dynamics and leads to a reduced probe light power. In
previous experiments, where we exposed cold atoms inside a high-Q
ring cavity to the dissipative and diffusive forces of an optical
molasses, we observed a threshold  behavior of the pump power
\cite{Cube04}. Only if the pump power was above threshold, the
CARL dynamics could be observed. In the present setup the atoms
are not exposed to a strong dissipative reservoir, such that it is
not fully clear, whether or not CARL with Bose-Einstein
condensates involves a threshold. The only channel to dissipation
in the present setup is the transmission through the cavity
mirrors, which is on the order of few ppm. By that the cavity
modes are coupled to the electromagnetic field of the
surroundings, which at room temperature is in a very good
approximation a zero-temperature reservoir of photons. The
interaction with this reservoir is therefore only dissipative, but
not diffusive. On the other hand Bose-Einstein-condensates are
object to quantum fluctuations. It is an open question, whether
such fluctuations can lead to a threshold behavior.\\

In the measurements shown in Fig. \ref{fig:pumppower} the pump
power could for technical reasons only be reduced to values of
$P_{+\textrm{,min}}\approx10^{-2}~\textrm{W}$ inside the cavity.
Within this accessible range, we observed that temperature effects
can lead to a strong threshold-like reduction of the probe light
power. The data agree very well with simulations (solid line)
using the experimentally measured parameters and a temperature of
the atoms of $T=800~\textrm{nK}$. The dotted line is a simulation
with the same parameters, but with a temperature of $T=0$. Above a
pump power of about $P_+=0.1~\textrm{W}$, both curves coincide.
Below this value the probe light power decreases much faster for a
finite temperature of the atoms. This behavior can be explained by
the fact, that depending on the temperature of the atoms the
initial momenta are spread over a number of momentum states given
by the Doppler width. However, only those atoms can participate in
the CARL dynamics, whose momenta are within the gain bandwidth,
which depends on the pump power. If by reduction of the pump power
the gain bandwidth gets smaller than the Doppler width, the
effective number of atoms for CARL is reduced. This leads to the
observed decrease of the probe light power. Another interesting
observable is the time difference $\Delta t_{1,2}$ between
successive maxima in the probe light power, because is represents
the typical time scale in which the momentum population is
shuffled between neighboring momentum states. The larger the gain
is, the faster is this time scale. This behavior can be seen in
Fig. \ref{fig:pumppower} (b), where our measurements agree well
with a simulation with the above given parameters and a
temperature of $T=0$. At finite temperature and small pump powers
a slight reduction of the time difference is observed, however
this correction is very small, such that for visibility we do not
plot the simulation with $T=800~\textrm{nK}$.

\subsection{Finesse and atom number}
\begin{figure}\label{fig:ndep}
            \includegraphics[height=5cm]{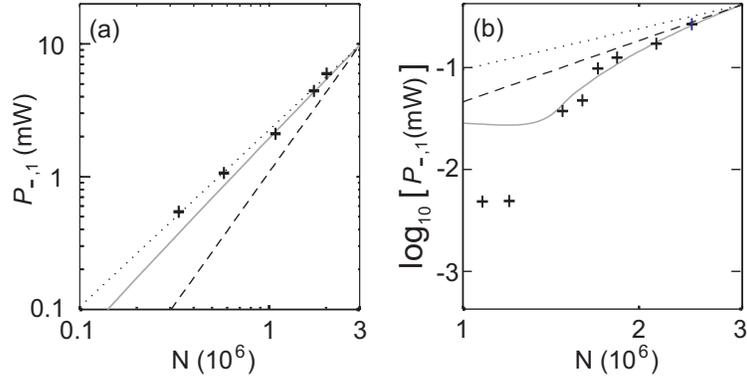}
            \caption{Measured dependency of the maximum probe light power on the atom number for two different
            values of the finesse. The experimental parameters are (a) $P_+=1.4~\textrm{W}$,
            $\lambda=796.1~\textrm{nm}$, $F=87000$ and (b) $P_+=66~\textrm{mW}$,
            $\lambda=795.4~\textrm{nm}$, $F=6400$. By comparison of the data points with the asymptotic
            behavior shown by the dotted and dashed lines, (a) can be identified as good-cavity regime and (b) as
            bad-cavity regime. The behavior is confirmed by measurements without free parameters (solid lines). The values
            of the data points are scaled by (a) 0.75 and (b) 2.8 in order to improve agreement with the simulations. We
            justify this multiplication by systematic errors which occur due to uncertainties in the calibration of the probe
            light power. Because the calibration depends on the polarization of probe light field, we assume a different
            scaling for low and good finesse. We emphasize that the relevant value, i.e. the dependency on the
            atom number, which is given by the slope in the logarithmic plot, is not changed by this pure multiplication.}
            \end{figure}

As explained above, the CARL model includes different regimes,
which are known as good-cavity and bad-cavity regime. With our
experiment we could for the first time explore both regimes by
changing the finesse of the cavity. Which regime is reached can be
analyzed from the way the probe light depends on the experimental
parameters. In the good-cavity regime, which is typical for CARL,
the maximum probe power scales like $P_{-,1}\propto
N^{4/3}P_+^{1/3}$, whereas in the bad-cavity regime the scaling is
like $P_{-,1}\propto N^{2}P_+$, which is typical for
superradiance. It is clear that at some point there must be a
transition from the one to the other behavior. This transition
however is not a sudden change, but rather a gradual crossover.
The above given scalings are reached in the limiting cases of a
very good and very bad cavity.\\
In order to determine the dominating regimes in our measurements,
we have been analyzing the dependence on atom number. The results
for two different values of the finesse are shown in Fig.
\ref{fig:ndep}. We compare our measurements with both limiting
scalings (dotted and dashed lines) and show simulations of the
CARL equations (solid lines) for the experimental parameters. The
simulations confirm that, although our measurements are still
close to the interface of both regimes, the expected behavior for
$F=87000$ (high finesse) is clearly CARL-like, whereas for
$F=6400$ (low finesse) it is clearly superradiance-like. In fact
our measurements show the expected behavior, i.e. for high finesse
we observe a scaling of the maximum probe power typical for the
good-cavity regime, whereas for low finesse it approximates the
bad-cavity behavior for large atom numbers. For low atom numbers
the CARL-dynamics is suppressed by mirror backscattering, which
can also be seen in the simulation, where this perturbation was
taken into account. This effect is discussed in \cite{Slama07b,
Slama07c}.

\subsection{Temperature}
\begin{figure}\label{fig:tdep}
            \includegraphics[height=7cm]{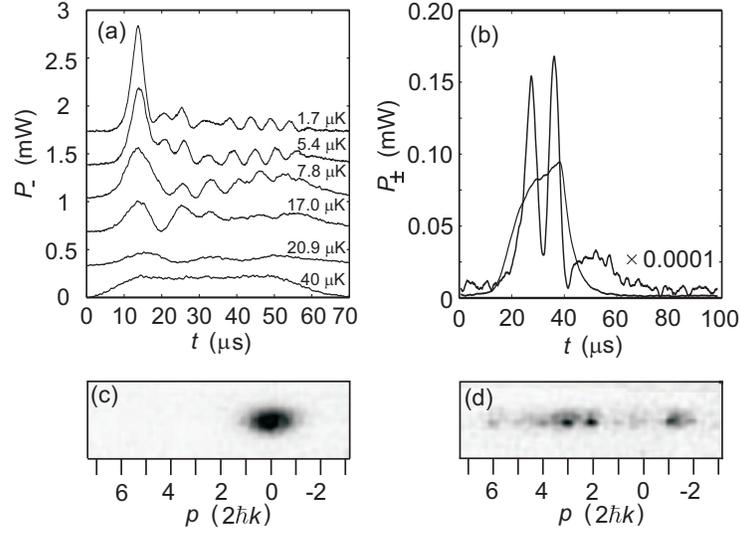}
            \caption{(a) Measured time signal of the probe light power for different atomic
            temperatures. For clarity the curves are shifted by
            0.35~mW from each other. Throughout all measurements the experimental parameters
            were held constant at $N=10^6$, $P_+=0.7~\textrm{W}$,
            $\lambda=796.1~\textrm{nm}$, $F=87000$. (b) Time
            signal of pump and probe light power during the CARL dynamics of a
            Bose-Einstein condensed cloud with experimental parameters $N\approx 10^5$, $P_+=0.7~\textrm{W}$,
            $\lambda=796.1~\textrm{nm}$, $F=87000$. The pump power
            is scaled down by a factor of $10^{-4}$ for clarity.
            The respective momentum distribution is shown in the time of
            flight pictures (c) before and (d) after the CARL
            dynamics, which have been recorded after $10~\textrm{ms}$ of adiabatic
            expansion.}
            \end{figure}

With our setup we can vary the temperature of the atoms in a range
from $T<1~\mu\textrm{K}$ to several tens of $\mu\textrm{K}$. This
allows us to systematically investigate the influence of the
atomic temperature on the CARL dynamics and identify the role of
quantum statistical effects ocurring with Bose-Einstein condensed
clouds. Figure \ref{fig:tdep} (a) shows recorded time signals of
the probe light power for thermal atoms with different
temperatures. The corresponding curves are vertically shifted for
clarity. We observe a decrease of the maximum probe power with
increasing temperature.\footnote{The bottom curve in Fig.
\ref{fig:tdep} shows no feature of CARL, but the signal is due to
mirror backscattering.} The reason is again, that the increasing
Doppler width of the atomic momentum distribution gets larger than
the gain bandwidth, which in this experiment had been kept
constant. This leads to a decreasing effective number of atoms
participating in the CARL dynamics. This also explains, why in
Fig. \ref{fig:tdep} (a) CARL is not observable at a temperature of
$T=40~\mu\textrm{K}$, while in our former experiments
\cite{Kruse03, Cube04} CARL was observed even with
$100~\mu\textrm{K}$ hot atoms. The reason lies in the difference
of the atom number of now $N=10^6$ to formerly $N=10^7$. A larger
atom number increases the gain bandwidth, which in turn increases
the possible Doppler width where CARL is observable. The important
point in Fig. \ref{fig:tdep} (a) is the fact that we do observe
CARL at temperatures which are far above the critical temperature
for Bose-Einstein condensation, which is below one
$\mu\textrm{K}$. This proves clearly that quantum statistical
effects are not important for the occurrence of both CARL and
SRyS, which, as shown above, are based on the same physical
phenomenon. Nevertheless it is interesting to see, what happens,
when CARL is done with a condensed cloud of atoms. This is
illustrated in Fig. \ref{fig:tdep} (b), which shows the measured
time signals of pump and probe power. The probe light power cannot
be directly compared with the one in Fig. \ref{fig:tdep} (a),
because the atom number in (b) is on the order of $N=10^5$
compared to $N=10^6$ in (a). The signal is for this reason weaker
in the case of a BEC. Apart from that quantitative difference,
however, we do not observe a qualitative change of the behavior.
This is also illustrated in Fig. \ref{fig:tdep} (c) and (d), which
show the momentum distribution corresponding to Fig. (b) in time
of flight pictures before and after the CARL dynamics. We observe
that, as expected for a BEC, initially a single momentum state is
occupied, whereas CARL leads to the population of several momentum
states. This broadening is typical for the semiclassical regime
and is equivalent to the observation of momentum spread in
\cite{Schneble03}.

\section{Conclusion}
The topic of our experiments presented in this paper is the
collective behavior of atoms inside a high-Q ring cavity, which is
known as CARL and leads to the self-organization of the atoms into
a periodic density grating. We have been able to measure CARL with
ultracold thermal and Bose-Einstein condensed atoms, at the same
time providing the first experimental realization of a BEC inside
an optical cavity. With this setup we could observe CARL in two
different regimes, which are known as bad-cavity and as
good-cavity regime \cite{Piovella97}. In these regimes the
dynamics is superradiance-like resp. CARL-like. The observed
characteristics of the instability allowed us to clearly identify
the two regimes and show the intrinsic connection between CARL and
superradiant Rayleigh scattering \cite{Inouye99}. By our former
observation of CARL with $T\approx 100~\mu\textrm{K}$ cold atoms
and by our new detailed analysis of CARL as a function of the
temperature between values of $T\approx 1~\mu\textrm{K}$ and
$T\approx 40~\mu\textrm{K}$ we proved experimentally that CARL and
hence also SRyS do not require a quantum degeneracy of the atoms,
but rely only on the cooperative behavior of the atoms \cite{Moore01}.\\
Although the quantum statistics of the atoms does not play a role
in our present experiments in future experiments it will be
interesting to extend our studies to the quantum regime, where the
dynamics is completely coherent. Within this regime photonic and
matter-waves may be coherently coupled during the CARL dynamics.
This may lead to the generation of entangled states between atoms
and photons \cite{Piovella03}.

\begin{theacknowledgments}
The work has been supported by the Deutsche Forschungsgemeinschaft
(DFG) under Contract No. Co 229/3-1. We like to thank Wolfgang
Ketterle, Nicola Piovella and Gordon Robb for helpful discussions.
\end{theacknowledgments}

\bibliographystyle{aipproc}   % if natbib is available

\end{document}